\documentclass[showpacs,aps,amssymb,floatfix,prd,amsmath,preprintnumbers]{revtex4}
\RequirePackage{float}
\usepackage{epstopdf}
\usepackage{hyperref}
\usepackage{capt-of}
\usepackage{graphicx}  
\usepackage{dcolumn}   
\usepackage{bm}
\begin{document}

\title{\bf  Revisiting The Coincidence Problem in $f(R)$ Gravitation}

\author{
P.K. Sahoo\footnote{ Department of Mathematics, Birla Institute of
Technology and Science-Pilani, Hyderabad Campus, Hyderabad-500078,
India,  Email:  pksahoo@hyderabad.bits-pilani.ac.in}, 
S. Bhattacharjee\footnote{Department of Astronomy, Osmania University, Hyderabad-500007,
India,  Email: snehasish.bhattacharjee.666@gmail.com}
}

\affiliation{ }

\begin{abstract}
The energy densities of dark matter (DM) and dark energy (DE) are of the same order at the present epoch despite the fact that both these quantities have contrasting characteristics and are presumed to have evolved distinctively with cosmic evolution. This is a major issue in standard $\Lambda$CDM cosmology and is termed "The Coincidence Problem" which hitherto cannot be explained by any fundamental theory. In this spirit, Bisabr \cite{bisabr} reported a cosmological scenario in $f(R)$ gravity where DM and DE interact and exchange energy with each other and therefore evolve dependently. We investigate the efficiency and model independancy of the technique reported in Bisabr \cite{bisabr} in addressing the Coincidence problem with the help of two $f(R)$ gravity models with model parameters constrained from various observations. Our result confirm the idea that not all scalar-tensor gravity theories and models can circumvent the Coincidence Problem and any cosmological scenario with interacting fluids is highly model dependent and hence alternate model independent theories and ideas should be nominated to solve this mystery.\\


\end{abstract}


\maketitle
  
\section{Introduction}

Dark energy is a mysterious form of energy permeating all of space with an anti-gravity effect fueling the acceleration of the 
universe \cite{1,100,99}. This enigmatic entity is the most dominant 
form of energy with a near 70\% share of the total energy
budget of the universe \cite{2}. The most popular model for dark energy is the cosmological constant ``$\Lambda$" originally postulated by Einstein \cite{gr} which he later regretted.
 
However hitherto no forms of dark energy have been observed which ushered many authors to modify the geometrical part of the field equations keeping the matter sector unchanged. Some of the well received modified gravity theories are $f(R)$ gravity by Buchdahl \cite{17b}, $f(T)$ gravity by Femaro \cite{16b}, $f(R,T)$ gravity by Harko \cite{7b} and $f(G)$ gravity by Nojiri \cite{22b}.

$f(R)$ gravity gained mainstream polularity following pioneering works by Starobinsky in modeling inflationary scenarios \cite{fr1}. Several $f(R)$ gravity models are known to explain the accelerated expansion of the universe \cite{fr2}. Viability of $f(R)$ gravity models have been derived in \cite{fr3}. Solar System experiments impose severe weak field constraints on $f(R)$ gravity thus ostracizing most of the models published heretofore \cite{fr4,fr41,fr42,fr43,fr44,fr45,fr46}, nonetheless some models still hold their feasibility \cite{fr5,fr51,fr52,fr53,fr54,fr55,fr56,fr57}. Readers are encouraged to see \cite{fr6,fr61,fr62,fr63} for recent reviews on $f(R)$ gravity.

Cosmology is muddled by a good deal of problems owing to the concealed nature of the cosmological constant. One of them being the so called "coincidence problem" which is concerned with the energy densities of dark matter ($\rho_{m}$) and dark energy ($\rho_{\varphi}$). Interestingly the ratio of their energy densities are of the same order at the present epoch regardless of the fact that both of these quantities evolve distinctively due to the acceleration of the universe. Adjacent to the possibility that the present epoch may be an inert regime at which the ratio of the two energy densities are constant, it is also possible that we live in a transient epoch at which the ratio varies slowly with respect to the expansion of the universe and hence arises the problem of coincidence \cite{bisabr}.

Many authors considered an unknown type of interaction between dark matter and dark energy which can lead to scale dependent evolution of their energy densities \cite{7,98,97,8,96,95,94}. The coincidence problem was addressed using such interactions by Bisabr \cite{bisabr} in $f(R)$ gravity, Rudra \cite{c1,c3,c2} in $f(T)$ gravity, $f(G)$ gravity and in $f(R,T)$ gravity. In this paper, we present two extensively studied $f(R)$ gravity models of the form: (a) $f(R) = R - \theta/R^{\zeta}$ and (b) $f(R) = R + \beta \log R - \alpha$ with model parameters ($\theta, \zeta, \beta, \alpha$) constrained from various observations in establishing the idea that any cosmological scenario with interacting fluids is highly model dependent. \\
The paper is organized as follows: In Section II we provide an overview of $f(R)$ gravity. In Section III we summarize the cosmological scenario reported by \cite{bisabr} in circumventing the coincidence problem. In Section IV we present two extensively studied $f(R)$ gravity models and report the in-efficiency and model dependency of the framework reported in \cite{bisabr} for a scale dependent evolution of DM and DE. In Section V we present our results and conclude the work.

\section{Theory}

The action in $f(R)$ gravity for a Jordan frame reads
\begin{equation}\label{1}
S=\frac{1}{16\pi G}\int \left[ \sqrt{-g} f(R)  + S_{m}\left( g_{\mu\nu},\psi\right) \right] d^{4}x
\end{equation}
where $G$ is the universal gravitational constant. We shall assume $G=1$. $g$ is the determinant of metric $g_{\mu\nu}$ and $S_{m}$ represents the action of non-baryonic dark matter which is a function of $g_{\mu\nu}$ and a scalar field $\psi$. According to Dolgov-Kawasaki instability \cite{dk} there exists some serious constraints on the forms of acceptable $f(R)$ gravity models. Such constraints demand a positive first and second order derivatives of $f(R)$ with respect to $R$. The positivity of the first order derivative fortify that scalar degree of freedom is not tachyonic while positivity of the second derivative establishes the fact that graviton is not a ghost particle \cite{bisabr}.\\
Due to equivalence of $f(R)$ gravity to being a scalar field minimally coupled to gravity with an appropriate potential function, we introduce a new set of variables \cite{bisabr} 
\begin{equation}\label{2}
\bar{g}_{\mu\nu}= \Phi g_{\mu\nu}
\end{equation}
\begin{equation}\label{3}
\varphi = \frac{1}{2\beta \sqrt{8\pi}}ln \Phi
\end{equation}
where $\Phi = df/dR = f^{'}_{R}$ and $\beta = \sqrt{1/6}$. This being a conformal transformation transforms the action into the following action in Einstein frame \cite{9,12,93} 
\begin{equation}\label{4}
S_{EF} = \frac{1}{2}\int\sqrt{-\bar{g}}\left[ \frac{1}{8\pi}\bar{R}  - 2V(\varphi) - \bar{g}^{\mu\nu}\bigtriangledown_{\mu \varphi}  \bigtriangledown_{\nu \varphi}   +  S_{m}\left(\bar{g}_{\mu\nu}e^{2\beta\sqrt{8\pi \varphi}}, \psi \right) \right]d^{4}x
\end{equation}
The scalar field $\varphi$ in Einstein frame reads 
\begin{equation}\label{5}
V(\varphi (R)) = \frac{R f^{'}(R)  -   f(R)}{16 \pi f^{'2}(R)}
\end{equation}
The conformal transformation generates a coupling between $\varphi$ with matter with identical strength $\beta = \sqrt{1/6}$ for all matter fields \cite{bisabr}. Variation with $\bar{g}^{\mu\nu}$ yields 
\begin{equation}\label{6}
\bar{G}_{\mu\nu} = 8 \pi \left( \bar{T}^{\varphi}_{\mu\nu} + \bar{T}^{m}_{\mu\nu} \right) 
\end{equation}
where 
\begin{equation}\label{7}
\bar{T}^{\varphi}_{\mu\nu} = \bigtriangledown _{\mu}\varphi\bigtriangledown _{\nu}\varphi - V(\varphi)\bar{g}_{\mu\nu} - \frac{1}{2}\bar{g}_{\mu\nu} \bigtriangledown^{\gamma}\varphi\bigtriangledown_{\gamma}\varphi
\end{equation}
\begin{equation}\label{8}
\bar{T}^{m}_{\mu\nu} = \frac{-2\delta S_{m} \left( \bar{g}_{\mu\nu} , \psi\right) }{\sqrt{\bar{g}}\delta  \bar{g}_{\mu\nu} }
\end{equation}
are energy-momentum tensors of scalar field and matter. The trace of equation (\ref{6}) is obtained as
\begin{equation} \label{9}
       \bar{T}^{m} = \bigtriangledown^{\gamma}\varphi\bigtriangledown_{\gamma}\varphi   -  \frac{\bar{R}}{8\pi} +  4V(\varphi) 
\end{equation}
Variation of action (\ref{4}) with respect to $\varphi$ yields
\begin{equation}\label{10}
-\beta \sqrt{8\pi } \bar{T}^{m} = \square \varphi - \frac{d V (\varphi)}{d \varphi}
\end{equation}
We note that $ \bar{T}^{\varphi}_{\mu\nu}$ \& $ \bar{T}^{m}_{\mu\nu}$ are not conserved separately but rather satisfy 
\begin{equation}\label{11}
\beta \sqrt{8\pi } \bigtriangledown_{\nu} \varphi \bar{T}^{m} = \bar{\bigtriangledown}^{\mu}\bar{T}^{m}_{\mu\nu}  =  - \bar{\bigtriangledown}^{\mu}\bar{T}^{\varphi}_{\mu\nu}
\end{equation}
We now consider a spatially homogeneous flat space-time described by FLRW metric of the form 
\begin{equation}\label{12}
ds^{2}= -dt^{2} + a^{2}(t)\left[dx^{2} + dy^{2} + dz^{2} \right] 
\end{equation}  
where $a(t)$ represents scale factor. Before applying the field equations to this metric we consider $\bar{T}^{m}_{\mu\nu}$ and $\bar{T}^{\varphi}_{\mu\nu}$ to be the energy momentum tensors of a perfect pressure less fluid with energy density $\bar{\rho}_{m}$ and a perfect fluid with pressure $p_{\varphi}= 1/2 \dot{\varphi}^{2} - V(\varphi)$ and energy density $\rho_{\varphi} =  1/2\dot{\varphi}^{2} + V(\varphi)$ respectively. Finally equations (\ref{6}) \& (\ref{10}) take the form 
\begin{equation}\label{13}
3H^{2} =  8 \pi \left( \rho_{\varphi} + \rho_{m}\right) 
\end{equation}
\begin{equation}\label{14}
2\dot{H} +3H^{2} = -8\pi\rho_{\varphi}\omega_{\varphi}
\end{equation}
\begin{equation}\label{15}
\ddot{\varphi} + \frac{dV(\varphi)}{d\varphi}  +  3H\dot{\varphi} =  -\beta\sqrt{8\pi} \rho_{m}
\end{equation}
where overhead dots represent time derivative and $\omega_{\varphi}=p_{\varphi}/\rho_{\varphi}$ represents equation of state (EoS) parameter of scalar field $\varphi$. Conservation equation (\ref{11}) and trace of (\ref{9}) give 
\begin{equation}\label{16}
\rho_{m}=\frac{R}{8\pi} + \dot{\varphi}^{2} - 4V(\varphi)
\end{equation}
\begin{equation}\label{17}
3H\rho_{m} + \dot{\rho}_{m} = B
\end{equation}
\begin{equation}\label{18}
3H\left(1 + \omega_{\varphi} \right) + \dot{\rho}_{\varphi} = -B
\end{equation}
where 

\begin{equation}\label{19}
B = \beta \sqrt{8\pi}\dot{\varphi}\rho_{m}
\end{equation}
represents the interaction term which vanishes when $\varphi = constant$. This happens when $f(R)$ is linearly dependent on $R$ according to equation (\ref{3}). The sign of $\dot{\varphi}$ or $B$ also governs the direction of transfer of energy. For $\dot{\varphi}< 0$, the energy gets transferred from dark matter to dark energy and reverse happens when $\dot{\varphi}> 0$ \cite{bisabr}.

\section{Solution to Coincidence Problem through interacting fluid}
The coincidence of similar energy densities between dark matter and dark energy is one of the cardinal hallmarks of cosmological constant problem \cite{14,92}. Different models have been proposed to suffice this observation by considering an interaction between these two components \cite{7,8} which lead to a net energy transfer between them and thus they are not conserved separately. As a result of which, the relative energy densities scale in the same way at all epoch. In \cite{bisabr} such an interaction was assumed and the model parameters of two phenomenological $f(R)$ gravity models were adjusted to fit the observations.\\
The ratio of energy densities are defined as \cite{bisabr} 
\begin{equation}
r\equiv \rho_{m}/ \rho_{\varphi}
\end{equation} 
and as reported in  \cite{bisabr}, $r$ should remain constant at all times.

Differentiating $r$ with time, we obtain 
\begin{equation}\label{20}
\dot{r}= \frac{\dot{\rho}_{m}}{\rho_{\varphi}}  -  r\frac{\dot{\rho}_{\varphi}}{\rho_{\varphi}}
\end{equation}
From equations (\ref{17}), (\ref{18}) \& (\ref{19}) we obtain 
\begin{equation}\label{21}
\dot{r} = \beta\sqrt{8\pi}\dot{\varphi}r(r+1) + 3H\omega_{\varphi}r
\end{equation}
Time derivative of Hubble parameter can be expressed as 
\begin{equation} \label{22}
\dot{H}  =  -H^{2}(1 + q)
\end{equation}
Substituting this to equation \ref{14} yields
\begin{equation} \label{23}
\omega_{\varphi} = \frac{H^{2} (1 - 2q)}{\sqrt{8 \pi}\rho_{\varphi}}
\end{equation}
Finally by substituting equations \ref{13}, \ref{23} into \ref{21} leads to 
\begin{equation}\label{24}
\dot{r} = rH(r + 1)(2q -1) + \beta\sqrt{8\pi}  (r+1)r
\end{equation}
Combining equations (\ref{5}), (\ref{13}) \& trace of (\ref{16}) gives 
\begin{equation} \label{25}
\dot{\varphi}^{2} = \frac{1}{8\pi}\left[3H^{2} (2q - 3) (1 - \frac{2}{f^{'}}) + \frac{3H^{2} r}{r + 1}  -2\frac{f}{f^{'2}}\right] 
\end{equation}
Plugging this to equation (\ref{24}) yields an expression which relate $\dot{r}$ with parameters $H$, $q$ and $r$. For the universe to advance towards a stage of constant $r$ or varying slowly with time than the scale factor, boils down to the following equation 
\begin{equation} \label{26}
M \left( H; r_{s}; f^{'}; q\right) = 0
\end{equation}
where 
\begin{equation}\label{27}
M \left( H; r_{s}; f^{'}; q\right) \equiv \sqrt{(r_{s} + 1)(2q -1)r_{s} + (1 + r_{s})\beta r_{s}\left[3(2q  - 3)(1 - \frac{2}{f^{'}}) + \frac{3 r_{s}}{1 + r_{s}} - \frac{2f}{f^{'2}H^{2}} \right] }
\end{equation}
and $r_{s}$ represents stationary value of $r$.\\
We note that at late-times signalized by $z = z_{*}$ we must obtain $\dot{r} = 0$ and denote $r_{s} = r_{*}$. Thus re-writing equation (\ref{26}) as 
\begin{equation}\label{28}
M\left(H_{*}; r_{*}; f_{*}; q_{*} \right) = 0 
\end{equation}
where 
\begin{equation}\label{29}
M\left(H_{*}; r_{*}; f_{*}; q_{*} \right) \equiv r_{*}\left(r_{*} + 1 \right) \left( 2q_{*} - 1\right) +\beta r_{*} (1 + r_{*} )\left[ 3 \left( 2q_{*} - 3 \right)\left( 1 - \frac{2}{f^{'}_{*}}\right) + \frac{3 r_{*}}{r_{*} + 1} - 2 \frac{f_{*}}{H_{*}^{2}f^{'2}_{*}}\right]^{\frac{1}{2}} 
\end{equation}
Here, $f_{*}$, $f_{*}^{'}$ represent late-time configurations of $f(R)$ and $f^{'}(R)$ which we obtain by substituting $R$ as
\begin{equation}\label{30}
R = 6H^{2}(1 - q)
\end{equation} 
at the redshift $z_{*}$. $f(R)$ gravity models characterizing $q(z)$ is given by\cite{15,16}
\begin{equation}\label{31}
q(z) =  \frac{q_{1} z + q_{2}}{(z + 1)^{2}} + \frac{1}{2}
\end{equation}
Fitting this model to Gold data set yields $q_{1} = 1.47$ \& $q_{2} = -1.46$ \cite{16} and $z_{*}=0.25$ \cite{bisabr}. Plugging all these into equation (\ref{31}) gives $q_{*}\cong -0.2$. We take $r_{*} = 3/7$ \cite{17,91,90} and $H_{*}= 70$ according to recent observations \cite{ligo}.
\section{$f(R)$ gravity models}
We seek to investigate whether the Coincidence Problem can be resolved through interacting fluids which demands Equation \eqref{29} to be zero and this is achieved by adjusting the model parameters of any scalar-tensor gravity model.\\
We now present two $f(R)$ gravity models of the form: (a) $f(R) = R - \theta/R^{\zeta}$ and (b) $f(R) = R + \beta \log R - \alpha$.  These cosmological models presented here allow the existence of the three evolutionary phases of the universe (radiation, matter and dark energy dominated eras) \cite{t8}, and delineate the current accelerated expansion of the universe without employing dark energy.\\
It is reported that model (a) depicts a switch from a decelerated to an accelerated phase at around $z = 1$ \cite{t46, t47}. This is achieved by studying a time evolving EoS parameter in terms of $z$ and also from the study of dynamical characteristics of $q(z)$. Similar results were reported for model (b) in terms of $\omega(z)$ and $q(z)$ \cite{t46,t47}.
\subsection{Power law gravity ($f(R) = R- \theta/R^{\zeta}$)}
Many authors \cite{t49,t50,t51,t52,t53} conducted both theoretical and observational studies on the viabilities of power-law $f(R)$ cosmological models in Palatini formalism. For the model $f(R) = R- \theta/R^{\zeta}$, $\zeta$ is a dimensionless parameter and $\theta \mathcal{O} H_{0}^{2(\zeta - 1)}$. Utilizing dynamical analysis, power-law $f(R)$ models were shown to depict the correct chronological evolution of the universe for $\zeta > -1$ and $\theta > 0$ \cite{t8}. In \cite{t9}, a study was conducted on the evolution of density perturbations for model (a). Numerical studies of the model were reported in \cite{t54}.\\
Although constraints from different combinations of CMB data, SNIa and large-scale structure greatly reduced the viable parameter-space to a very small domain ($\theta\approx 10^{-3}$) favoring $\Lambda$CDM cosmological model \cite{t13,t42}, other studies integrating BAO and CMB data produced $(\zeta, \theta) = (0.027_{-0.23}^{+0.42}, 4.63^{+2.73}_{-10.6})$ \cite{t8}. These estimates were further upgraded with amalgamations of BAO, SNIa, CMB data, coupled with Hubble parameter evaluations \cite{t55,t57}, and strong lensing analysis \cite{t14,t58}. From the investigation of deceleration parameter and statefinder diagnostics, cosmographic constraints were reported in \cite{t48, t46}. Current constraints yielded from standard cosmological rulers in quasars narrowed the parameter ranges even further:  $(\zeta, \theta) = (0.052_{-0.071}^{+0.077}, 4.736^{+0.882}_{-0.681})$ \cite{t59}. 
\begin{figure}[H]
\centering
  \includegraphics[width=3.5in]{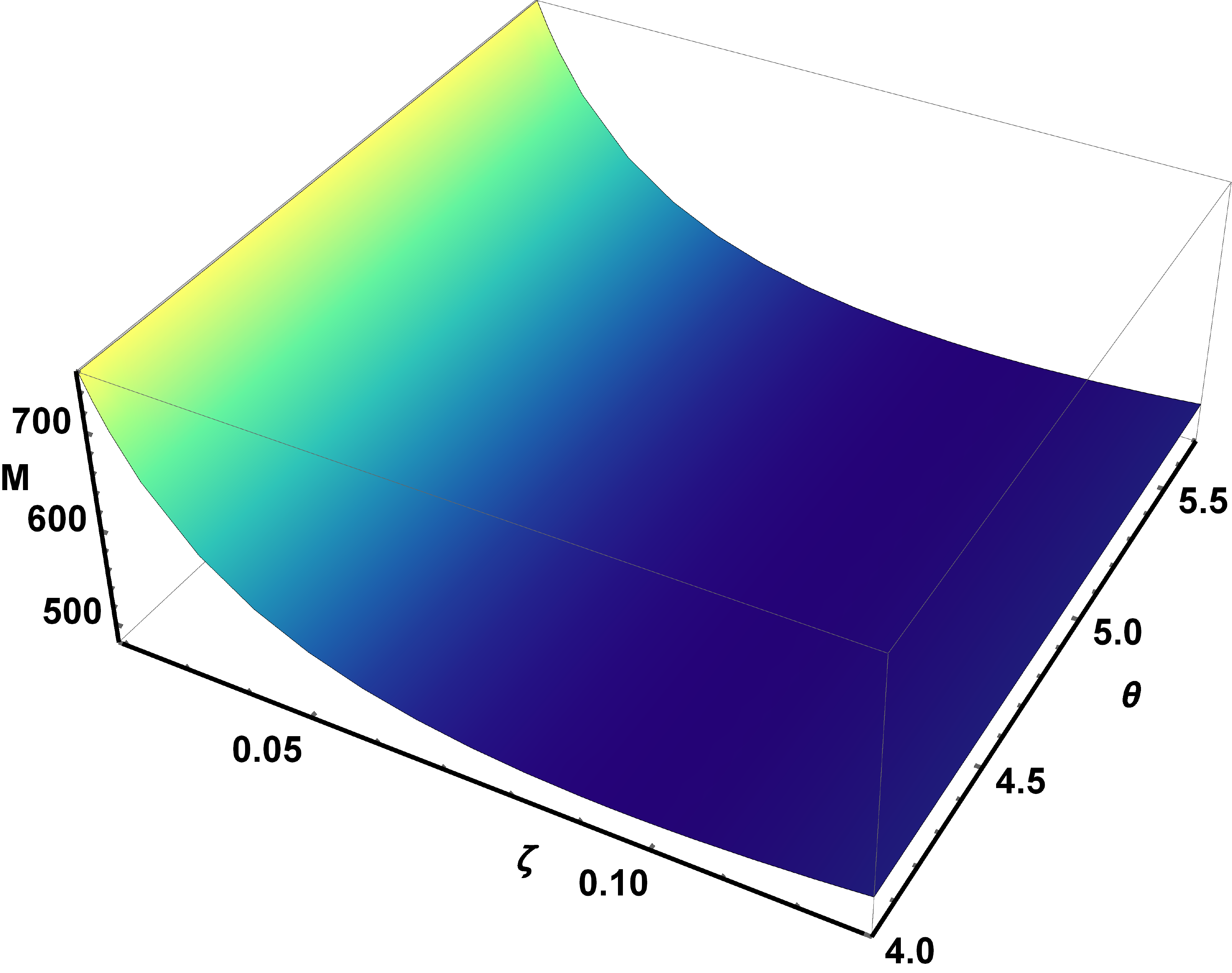}
  \caption{Plot of $M$ for the allowed parameters spaces of $\zeta$ and $\theta$}
  \label{f1}
\end{figure}
As reported in \cite{bisabr}, $M$ has to be nullified in order to have a scale dependent evolution of DM and DE at all times. In Figure \eqref{f1} we observe that the function $M$ does not attain a null value for the constrained parameter spaces of $\zeta$ and $\theta$ at the present epoch. Therefore a power law $f(R)$ gravity model does not support a scale dependent evolution of DE and DM.
\subsection{Logarithmic gravity ($f(R) = R + \beta \log R - \alpha$)}
Logarithmic $f(R)$ cosmological models were first studied in \cite{t60}, where logarithmic terms were employed to suffice the current accelerated expansion. At high redshifts, these models correspond to standard GR evolution. In \cite{t8}, logarithmic gravity models were reported to allow the presence of radiation, matter and dark energy dominated epochs. When compared to the power law model, constraints on model parameters for the logarithmic models enumerated in units of $H_{0}^{2}$, have not been extensively studied. Different datasets coming from BAO, CMB and SNIa were reported in \cite{t8} with $(\beta, \alpha) = (0.11^{+1.75}_{-1.11}, 4.62_{-5.58}^{+3.54})$. A more recent study reported $(\beta, \alpha) =( -0.48^{+1.26}_{-2.67}, 3.58)$ \cite{t46,t47}.\\
For this cosmological model, we observe $M$ to be a complex number for the allowed parameter spaces of $\beta$ and $\alpha$ and reads
\begin{equation}
M \approx -0.86 + 0.45 i
\end{equation} 
where, $i = \sqrt{-1}$. Thus, for both the models we are unable to obtain a null value of function $M$ which clearly is in disagreement with a cosmological scenario involving a scale dependent evolution of DM and DE.

\section{Conclusions}
The energy densities of dark matter and dark energy are of the same order at the present epoch despite the fact that both these quantities are completely different from each other and are presumed to have evolved distinctively over the entire cosmic aeon. Recent observations suggest that the equality between the energy densities of DM and DE took place at very low redshift of $z\sim0.55$ \cite{aspect}. This equality at low redshifts then implies that DE dominated era started recently which ultimately give rise to the Coincidence problem "why now?" \cite{aspect}. 
Due to the lack of any fundamental theory which can explain this "why now?" problem, researchers \cite{bisabr,7,98,97,8,96,95,94, c1,c3,c2} were inspired to presume that DE and DM evolve dependently by exchanging energy with each other. \\
The idea that such dark components may interact by exchanging energy, so their energy densities always remain of the same order is a very standard way to circumvent the so-called Coincidence Problem existing within the $\Lambda$CDM Cosmological Concordance Model. Such "tracking", or more elaborated scenarios, have been widely studied in the context of extended (also dubbed modified) theories of gravity, particularly special emphasis was historically put on scalar field models with either conformal or conformal plus disformal couplings.\\
Our results can be summarized as follows:
\begin{itemize}
\item We investigate the efficiency and model independancy of the technique reported in \cite{bisabr} in addressing the Coincidence problem against observational grounds with the assistance of two widely studied paradigmatic scalar-tensor $f(R)$ gravity models of the form $f(R) = R- \theta/R^{\zeta}$ and $f(R) = R + \beta \log R - \alpha$ with model parameters constrained from cosmological observations.
\item We found that for both the models, Equation \eqref{29}, does not become zero and this does not favor a cosmological scenario involving scale dependent evolution of dark components according to the approach reported in \cite{bisabr} . 
\item Our result confirm the idea that not all scalar-tensor gravity theories and models can circumvent the Coincidence Problem and any cosmological scenario with interacting fluids is highly model dependent.
\item Interestingly, It may \textbf{also} be quite reasonable to think that a scale dependent evolution of DM and DE may not take place after all and alternate model independent theories and ideas should be put forwarded to solve this conundrum.
\end{itemize}

\section*{Acknowledgments}

PKS acknowledge DST, New Delhi, India for providing facilities through DST-FIST lab, Department of Mathematics, BITS-Pilani, Hyderabad Campus where a part of the work was done.

\end{document}